\documentstyle[aps,twocolumn,epsfig]{revtex}
\begin{document}
\draft

\title{QCD prediction for heavy boson transverse momentum 
distributions}
\author{Jianwei Qiu and Xiaofei Zhang}
\address{Department of Physics and Astronomy,
         Iowa State University;
         Ames, Iowa 50011, USA}

\date{December 4, 2000}
\maketitle
\begin{abstract}
We investigate the predictive power of Collins, Soper, and
Sterman's $b$-space QCD resummation formalism for transverse momentum
($Q_T$) distributions of heavy boson production in  
hadronic collisions.  We show that the predictive power of the
resummation formalism has a strong dependence on the collision energy
$\sqrt{S}$ in addition to its well-known $Q^2$ dependence, and the
$\sqrt{S}$ dependence improves the predictive power at collider
energies.  We demonstrate that at Tevatron and the LHC
energies, the $Q_T$ distributions derived from $b$-space resummation
are not sensitive to the nonperturbative input at large
$b$, and give good descriptions of the $Q_T$
distributions of heavy boson production at all 
transverse momenta $Q_T \leq Q$.  
\end{abstract}



With new data from Fermilab Run II on the horizon and the LHC in the
near future, we expect to test Quantum Chromodynamics (QCD) to a new
level of accuracy, and also expect that a better understanding of
QCD will underpin precision tests of the Electroweak interactions and
particle searches beyond the Standard Model \cite{QCD-rpt}.  
As pointed out in Ref.~\cite{QCD-rpt}, the description of vector and
scalar boson production properties, in particular their transverse
momentum ($Q_T$) distribution, is likely to be one of the most
intensively investigated topics at both Fermilab and the LHC,
especially in the context of Higgs searches.  The main purpose of
this Letter is to quantitatively demonstrate the predictive power
of QCD resummation formalism for the $Q_T$ distributions of heavy
boson production at Tevatron and the LHC.  In particular, we
concentrate on the small transverse momentum region: $Q_T\leq Q$,
where the bulk of the data is.  This region is also most
relevant to the hadronic Higgs production.    

When $Q_T \ll Q$, the $Q_T$ distribution calculated
in conventional fixed-order perturbation theory
receives large logarithm, $\ln(Q^2/Q_T^2)$, at every power of
$\alpha_s$, which is a
direct consequence of emissions of soft and collinear gluons by 
incoming partons.  Therefore, at sufficiently small $Q_T$, the
convergence of conventional perturbative expansion in powers of
$\alpha_s$ is impaired, and the logarithms must be resummed.   

Resummation of the large logarithms can be carried out either in 
$Q_T$-space directly, or in the impact parameter, $b$-space, which is 
the Fourier conjugate of $Q_T$-space.  It was first shown by
Dokshitzer, Diakonov, and Troyan that in the double leading logarithm 
approximation, the dominant contributions in small $Q_T$ 
region can be resummed into a Sudakov form factor \cite{DDT-qt}.  
By imposing the transverse momentum conservation without assuming 
the strong ordering in transverse momenta of radiating gluons, 
Parisi and Petronzio introduced the $b$-space resummation method which 
allows to resum some subleading logarithms \cite{PP-b}.  
Using the renormalization group equation technique, 
Collins and Soper improved the $b$-space resummation to resum 
all logarithms as singular as, $\ln^m(Q^2/Q_T^2)/Q_T^2$, 
as $Q_T\rightarrow 0$ \cite{CS-b}.  Using this renormalization 
group improved $b$-space resummation, Collins, Soper, and 
Sterman (CSS) derived a formalism for the transverse momentum
distributions of vector boson production in hadronic collisions
\cite{CSS-W}.  This formalism, which is often called CSS formalism,  
can be also applied to the hadronic production of Higgs bosons
\cite{QCD-rpt}.  

For Drell-Yan vector boson ($V=\gamma^*,W^{\pm},Z$) production in
hadronic collisions between hadrons $A$ and $B$,
the CSS formalism has the following generic form \cite{CSS-W}  
\begin{eqnarray}
\frac{d\sigma_{A+B\rightarrow V+X}}{dQ^2\, dy\, dQ_T^2} &=&
\frac{1}{(2\pi)^2}\int d^2b\, e^{i\vec{Q}_T\cdot \vec{b}}\,
\tilde{W}(b,Q,x_A,x_B) 
\nonumber \\
&+& Y(Q_T,Q,x_A,x_B)\, ,
\label{css-gen}
\end{eqnarray}
where $\tilde{W}$ term gives the dominant contribution when
$Q_T\ll Q$, while $Y$ term gives corrections that are negligible
for small $Q_T$, but become important when $Q_T\sim Q$.  In
Eq.~(\ref{css-gen}), $x_A= e^y\, Q/\sqrt{S}$ and 
$x_B= e^{-y}\, Q/\sqrt{S}$ with the rapidity $y$ and collision
energy $\sqrt{S}$.  The $\tilde{W}$ in Eq.~(\ref{css-gen}) 
includes all powers of large logarithms from $\ln(1/b^2)$ to
$\ln(Q^2)$ and has the following form 
\cite{CSS-W} 
\begin{equation}
\tilde{W}(b,Q,x_A,x_B) = 
{\rm e}^{-S(b,Q)}\, \tilde{W}(b,c/b,x_A,x_B)\, ,
\label{css-W-sol}
\end{equation}
where $c$ is a constant of order one \cite{CSS-W,QZ2}, and
$S(b,Q) = \int_{c^2/b^2}^{Q^2}\, 
  \frac{d{\mu}^2}{{\mu}^2} \left[
  \ln\left(\frac{Q^2}{{\mu}^2}\right) 
  A(\alpha_s({\mu})) + B(\alpha_s({\mu})) \right],
$
with $A(\alpha_s)$ and $B(\alpha_s)$ perturbatively calculable
\cite{CSS-W}.  The $\tilde{W}(b,c/b,x_A,x_B)$ in 
Eq.~(\ref{css-W-sol}) depends on only one momentum scale, $1/b$, and
is perturbatively calculable as long as $1/b$ is large
enough.  The large logarithms from $\ln(c^2/b^2)$ to $\ln(Q^2)$ in
$\tilde{W}(b,Q,x_A,x_B)$ are completely resummed into the exponential
factor $\exp[-S(b,Q)]$.  

Since the perturbatively resummed $\tilde{W}(b,Q,x_A,x_B)$ in
Eq.~(\ref{css-W-sol}) is only reliable for the small $b$ region, an
extrapolation to the large $b$ region is necessary in order to
complete the Fourier transform in Eq.~(\ref{css-gen}).
In the CSS formalism, a variable $b_*$ 
and a nonperturbative function $F^{NP}(b,Q,x_A,x_B)$
were introduced \cite{CSS-W}, 
\begin{eqnarray}
\tilde{W}_{\rm CSS}(b,Q,x_A,x_B) &\equiv &
\tilde{W}(b_*,Q,x_A,x_B)
\nonumber \\
&\times & 
F^{NP}(b,Q,x_A,x_B) ,
\label{css-W-b}
\end{eqnarray}
where $b_*=b/\sqrt{1+(b/b_{max})^2} < b_{max} = 0.5$~GeV$^{-1}$, and
$F^{NP}$ has a Gaussian-like form, $F^{NP} \sim \exp(-\kappa
b^2)$ and the parameter $\kappa$ has some dependence on $Q^2$, 
$x_A$, and $x_B$.  The data are not inconsistent with
such a form \cite{Davis,AK,LY,Ellis-1}.  However, improvements are
definitely needed for the precision tests of the theory
\cite{QCD-rpt,Ellis-2}. 

Although the $b$-space resummation formalism has been successful in
interpreting existing data, it was argued \cite{QCD-rpt,Ellis-2}
that the formalism has many drawbacks associated with working in
impact parameter space.  As listed in Ref.~\cite{QCD-rpt}, the
first is the difficulty of matching the resummed and fixed-order
predictions; and the second is to know quantitative difference between
the prediction and the fitting because of the introduction of a
nonperturbative $F^{NP}$.  In viewing of these
difficulties, major efforts have been devoted to resum the large
logarithms directly in $Q_T$-space \cite{QCD-rpt,Ellis-2}.

In the following, we argue and demonstrate that both of these
drawbacks can be overcomed.  We show that 
the $b$-space formalism works smoothly for all $Q_T \leq Q$.
We demonstrate that the $Q^2$ and $\sqrt{S}$ dependence of the
resummed $b$-space distribution ensure that the Fourier transform is
completely dominated by the small $b$ region in 
high energy collisions, and consequently, the $Q_T$ distribution is
insensitive to the details of $F^{NP}$.  

It was known \cite{QCD-rpt} that the $b$-space resummed $Q_T$
distribution from Eq.~(\ref{css-gen}) becomes unphysical or negative
when $Q_T$ is large.  For example, a matching between the resummed and
fixed-order calculations has to take place at $Q_T \sim 50$~GeV for
$W$ production when these two predictions cross over \cite{Ellis-2}.
On the other hand, we expect that the 
predictions given by the $b$-space resummation in Eq.~(\ref{css-gen})
should work better when $Q_T$ is large, because the perturbatively
calculated $Y$ term dominates and the predictions should be less
sensitive to $\tilde{W}$ and its nonperturbative input.  We find
that this puzzle was mainly caused by the lack of numerical accuracy
of the Fourier transform from the $b$-space to $Q_T$-space.

Since there is no preferred transverse direction, the $\tilde{W}$ in
Eq.~(\ref{css-gen}) is a function of $b=|\vec{b}|$, and the Fourier
transform can be written as
\begin{eqnarray}
&\ &
\frac{1}{(2\pi)^2}\int d^2b\, e^{i\vec{Q}_T\cdot \vec{b}}\,
\tilde{W}(b,Q,x_A,x_B) 
\nonumber \\
&=& 
\frac{1}{2\pi} \int_0^\infty db\, b\, J_0(Q_T b)\, 
{\rm e}^{-S(b,Q)}\, \tilde{W}(b,\frac{c}{b},x_A,x_B)\, ,
\label{css-W-F}
\end{eqnarray}
where $J_0(z)$ with $z=Q_T b$ is the Bessel function.  Because of the
oscillatory nature of the Bessel function, high accuracy of the
numerical integration over $b$ is crucial for a reliable result.  The
number of oscillations strongly depends on the value of $Q_T$ for the
same range of $b$.  For example, when $b \in (0, 2)$~GeV$^{-1}$, 
$J_0(Q_T b)$ crosses zero 0, 6, and 63 times for $Q_T=1$, 10, and
100~GeV, respectively.  It is clear that numerical accuracy is
extremely important for the large $Q_T$ region.  
We noticed that most work
published in the literature used some kind of asymptotic form to
approximate the Bessel function when $z=Q_T b$ is large. We believe
that the asymptotic form is a source of the uncertainties observed for
the large $Q_T$ region.  Instead of using an asymptotic form, we use
an integral form for the Bessel function $J_0(z) = \frac{1}{\pi}\, 
\int_0^\pi \cos\left(z\sin(\theta)\right) d\theta$.
The great advantage of using an integral form is that we can control
the numerical accuracy of the Bessel function by improving the
accuracy of the integration.  With the integral form of the Bessel
function, we show below that the $b$-space resummed  
$Q_T$ distributions are smoothly consistent with data for all
transverse momenta up to $Q$ \cite{QZ2}.   

The predictive power of the CSS resummation formalism relies on the
fact that the integration over $b$ in Eq.~(\ref{css-W-F}) is dominated
by the region where $b\sim 1/Q$, because the exponential factor
$\exp[-S(b,Q)]$ in Eq.~(\ref{css-W-F}) suppresses the $b$-integral
when $b$ is larger than $1/Q$ \cite{CSS-W}.  Using the saddle point
method, it was shown 
\cite{PP-b,CSS-W} that even at $Q_T=0$, the $b$-integration in
Eq.~(\ref{css-W-F}) is dominated by an impact parameter of order
\begin{equation}
b_{\rm SP} = \frac{1}{\Lambda_{\rm QCD}}
  \left( \frac{\Lambda_{\rm QCD}}{Q} \right)^{\lambda}
\label{css-bsp}
\end{equation}
where $\lambda=16/(49-2n_f)\approx 0.41$ for quark flavors $n_f=5$.
When $Q\sim M_Z$ or $M_W$, the momentum scale $c/b_{\rm SP}$ in
Eq.~(\ref{css-W-F}) is of a few GeV, at which the perturbation theory
is expected to work.  Consequently, the
predictive power of the CSS formalism should not
be very sensitive to the $F^{NP}$ at large $b$ as long as $Q$ is
large.  

The $b_{\rm SP}$ in Eq.~(\ref{css-bsp}) was derived with an assumption
that the $b$-dependence in $\tilde{W}(b,c/b,x_A,x_B)$ is smooth around
$b_{\rm SP}$.  We find that the $b$-dependence in
$\tilde{W}(b,c/b,x_A,x_B)$ is important.  When $x_A$ and $x_B$ are
small, this $b$-dependence reduces the numerical value of saddle
point, and consequently, increases the predictive power of the
$b$-space resummation formalism \cite{QZ2}.

\begin{center}
\begin{figure}
\epsfig{figure=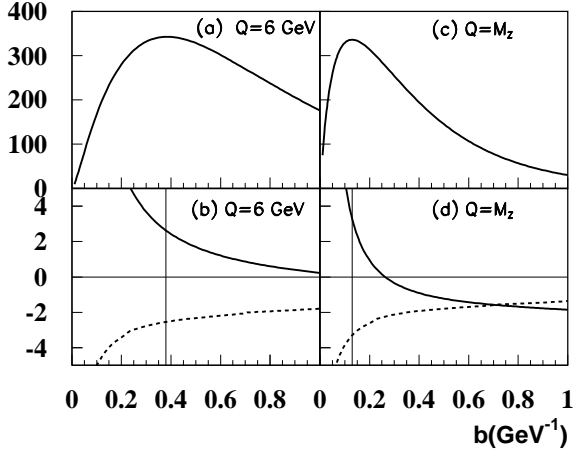,width=3.0in}
\caption{Integrand of the $b$-integration in
Eq.~(\protect\ref{css-W-F}) at $Q_T=0$ as a function of $b$ for
$Q=6$~GeV (a) and $Q=M_Z$ (c) with an arbitrary normalization; 
and the first (solid) and second (dashed) terms in
Eq.~(\protect\ref{saddle}) as a function of $b$ in (b) and (d) at the 
respective $Q$.}
\label{fig1}
\end{figure} 
\end{center}

Taking into account the full $b$-dependence of
$\tilde{W}(b,c/b,x_A,x_B)$, the saddle point for the $b$-integration
in Eq.~(\ref{css-W-F}) at $Q_T=0$ is determined by solving the
following equation 
\begin{equation}
 \frac{d}{db}\ln\left(b{\rm e}^{-S(b,Q)}\right)
+\frac{d}{db}\ln\left(\tilde{W}(b,\frac{c}{b},x_A,x_B)\right)
=0 .
\label{saddle}
\end{equation}
The $b_{\rm SP}$ in Eq.~(\ref{css-bsp}) corresponds to a solution of
Eq.~(\ref{saddle}) without the second term and keeps only the first
order term of the $A(\alpha_s)$ in $S(b,Q)$.  As shown in 
Ref.~\cite{QZ2}, the second term in Eq.~(\ref{saddle}) is proportional
to the evolution of parton distribution $\phi(x,\mu)$:
$-\frac{d}{d\ln(1/b)} \phi(x,\mu=\frac{c}{b})$.
The evolution $(d/d\ln\mu) \phi(x,\mu)$ is positive (or
negative) for $x<x_0\sim 0.1$ (or $x>x_0$), and is very steep when $x$
is far away from $x_0$.  Therefore, the second term in
Eq.~(\ref{saddle}) {\it reduces} the numerical value of the saddle
point when $x_A$ and $x_B$ are much smaller than $x_0$.
As a demonstration, let $Q=6$~GeV and $\sqrt{S}=1.8$~TeV.  
Using CTEQ4 parton distribution and 
$\Lambda_{\rm QCD}(n_f=5)=0.202$~GeV \cite{CTEQ4}, 
one derives from Eq.~(\ref{css-bsp}) 
that $b_{\rm SP}\approx 1.2$~GeV$^{-1}$, and might
conclude that perturbatively resummed prediction for $Q_T$
distribution at the given values of $Q$ and $\sqrt{S}$ is not
reliable.  However, as shown in Fig.~\ref{fig1}(a), the integrand of
the $b$-integration in Eq.~(\ref{css-W-F}) has a nice saddle point at
$b_0\approx 0.38$~GeV$^{-1}$, which is within the perturbative region.
This is due to the fact that $x_A\sim x_B\sim 0.003$ are very small
and the second term in Eq.~(\ref{saddle}) is negative and important.
Fig.~\ref{fig1}(b) shows that the second term 
in Eq.~(\ref{saddle}) (dashed line) cancels the first term (solid
line) at the saddle point $b_0$, which is much smaller than the 
$b_{\rm SP}$ estimated from Eq.~(\ref{css-bsp}).
In Fig.~\ref{fig1}(c) and \ref{fig1}(d), we show the effect of the 
second term in Eq.~(\ref{saddle}) on the saddle point of $Z$
production at the LHC energy.  At $\sqrt{S}=14$~TeV, the  
$\sqrt{S}$ dependence for $Z$ production improves the value of the
saddle point from $b_0=0.24$~GeV$^{-1}$ at $\sqrt{S}=1.8$~TeV to 
0.13~GeV$^{-1}$, in comparison to an estimated 
$b_{\rm SP} \approx 0.40$~GeV$^{-1}$ from Eq.~(\ref{css-bsp}).
The narrow width of the $b$-distribution shown in Fig.~\ref{fig1}(c)
also ensures that the $b$-integration is dominated by $b\sim b_0$.
Similarly, we find the same improvements on the saddle point
and width of the $b$-distribution for Higgs production at
LHC energy \cite{QZ2}.  In conclusion, the $b$-space resummation
formalism for heavy boson production at collider energies should not
be very sensitive to the nonperturbative input $F^{NP}$ at large $b$. 

To quantitatively demonstrate the sensitivities on the $F^{NP}$, we
reexam the extrapolation, $\tilde{W}_{\rm CSS}$, defined in
Eq.~(\ref{css-W-b}).  We find that using the fitting parameters from 
Refs.~\cite{Davis,AK,LY,Ellis-1} to fix the $F^{NP}$, the ratio, 
$\tilde{W}_{\rm CSS}(b,Q,x_A,x_B)/\tilde{W}(b,Q,x_A,x_B)$, differs
from one by as much as 20 percents within the perturbative region:
$b<b_{max}\sim 0.5$~GeV$^{-1}$.  That is, the $\tilde{W}_{\rm
CSS}(b,Q,x_A,x_B)$ introduces a significant fitting parameter
dependence to the resummed $b$-distribution in the perturbative
region. 

In order to separate the perturbative prediction in small $b$ region
from the nonperturbative physics at large $b$, we review the
resummation of large logarithms in the CSS formalism, and introduce a
new functional form for the extrapolation.  In Ref.~\cite{CSS-W}, the
large logarithms are resummed by solving the evolution equation
\begin{eqnarray}
\frac{\partial}{\partial\ln Q^2} &\ & {\hskip -0.11in}
\tilde{W}(b,Q,x_A,x_B)
= \left[ K(b\mu,\alpha_s(\mu)) \right.
\nonumber \\
&+& 
\left. G(Q/\mu,\alpha_s(\mu)) \right]
\tilde{W}(b,Q,x_A,x_B)\, ,
\label{css-W-evo}
\end{eqnarray}
from $\ln(c^2/b^2)$ to $\ln(Q^2)$, and the renormalization group
equations for $K$ and $G$ from $\ln(c^2/b^2)$ to $\ln(\mu^2)$ and from
$\ln(\mu^2)$ to $\ln(Q^2)$, respectively \cite{CSS-W}.  
Since the evolution equation and the renormalization
group equations do not include any power corrections, the solution,
$\tilde{W}$ in Eq.~(\ref{css-W-sol}), is valid 
only for $b<b_{c}$ with $\ln(1/b_{c}^2) \sim b_{c}^2$ (or $b_{c}\sim
0.75$~GeV$^{-1}$).  The choice of $b_{max}=0.5$~GeV$^{-1}$ in
Ref.~\cite{CSS-W} is consistent with the approximation.

Taking advantage of our early conclusion that heavy boson production
at collider energies should not be very sensitive to the large $b$
region, we extrapolate $\tilde{W}(b,Q,x_A,x_B)$ to the
large $b$ region without introducing the $b_*$ \cite{QZ2}.  
For $b<b_{max}$, the $\tilde{W}(b,Q,x_A,x_B)$ is the same as the
perturbatively calculated one given in Eq.~(\ref{css-W-sol}).  
For $b>b_{max}$, we solve the evolution equation in
Eq.~(\ref{css-W-evo}) from $\ln(c^2/b_{max}^2)$ to $\ln(Q^2)$, and
solve the renormalization group equations for $K$ and $G$ from
$\ln(c^2/b^2)$ to $\ln{\mu^2}$ and from $\ln(\mu^2)$ to $\ln(Q^2)$,
respectively.  Because $b > b_{max}$, 
we add possible power corrections to the renormalization group
equations of $K$ and $G$.  We find \cite{QZ2}
\begin{eqnarray}
\tilde{W}_{QZ}(b,Q,x_A,x_B) &=& \tilde{W}(b_{max},Q,x_A,x_B)
\nonumber \\
&\times &
F^{NP}(b,b_{max},Q,x_A,x_B),
\label{qz-W-sol}
\end{eqnarray}
for $b>b_{max}$.
Including only the first power correction, the $F^{NP}$ has the
following functional form \cite{QZ2} 
\begin{equation}
\ln(F^{NP}) = -g_1 \left[ (b^2)^\alpha - (b_{max}^2)^\alpha \right]
              -g_2 \left[ b^2 - b_{max}^2 \right] ,
\label{qz-fnp}
\end{equation}
where $g_1$, $g_2$ and $\alpha (<1)$ are parameters.  Their dependence
on $Q$, $x_A$, and $x_B$, which are more relevant for the low $Q^2$
Drell-Yan data, are explained in Ref.~\cite{QZ2}.  In
Eq.~(\ref{qz-fnp}), the first term corresponds to a direct
extrapolation of the logarithmic contributions to the function $K$ to
large $b$ region.  The $(b^2)^\alpha$ dependence is a result of
replacing a series of logarithmic dependence on $\mu^2$ in the
renormalization group equations by $(\mu^2)^\alpha$.
The second term is a consequence of the first power correction
($1/\mu^2$) to the renormalization group equations.  Since
the saddle point has a small numerical value in $b$, high power
corrections to the renormalization group equations of the $K$ and $G$,
which are sensitive to the very large $b$ region, could be 
neglected.  In addition, we neglect a term, 
$\ln(\tilde{W}(b,c/b_{max},x_A,x_B)) -
 \ln(\tilde{W}(b_{max},c/b_{max},x_A,x_B))$,
for the $\ln(F^{NP})$ by assuming that parton distributions are
saturated when $b>b_{max}$.  More detailed discussions on the
functional form of the $F^{NP}$ are given in Ref.~\cite{QZ2}.  

\begin{center}
\begin{figure}
\epsfig{figure=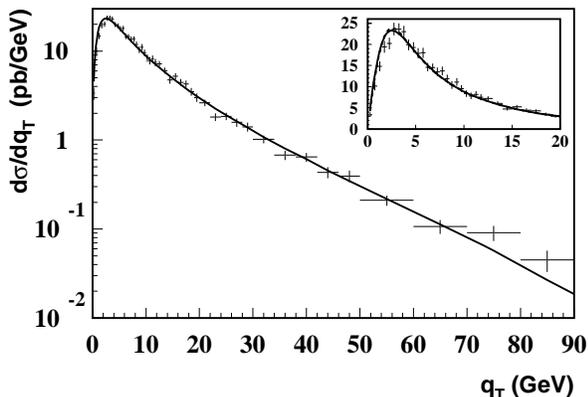,width=3.0in}
\caption{Comparison between the $b$-space resummed $Q_T$ distribution
and CDF data \protect\cite{CDF-Z}. The inset shows the $Q_T<20$~GeV
region.} 
\label{fig2}
\end{figure} 
\end{center}

\begin{center}
\begin{figure}
\epsfig{figure=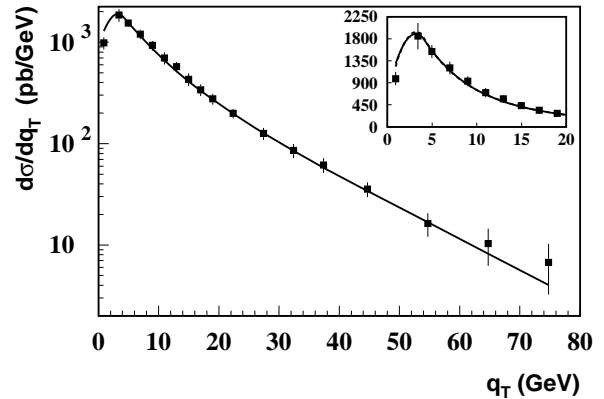,width=3.0in}
\caption{Comparison between the $b$-space resummed $Q_T$ distribution
and D0 data \protect\cite{D0-W}.  The inset shows the $Q_T<20$~GeV
region.}
\label{fig3}
\end{figure} 
\end{center}

We quantitatively test the sensitivities on the $F^{NP}$ by studying
the dependence on $b_{max}$, $g_2$, and $\alpha$.  We first set
$g_2=0$ (no power corrections) and fix $g_1$ and $\alpha$ in 
Eq.~(\ref{qz-fnp}) by requiring the first and second derivatives of
the $\tilde{W}$ to be continuous at $b=b_{max}=0.5$~GeV$^{-1}$.  We
plot our predictions (solid lines) to the $Q_T$ distributions of $Z$
and $W$ production at Tevatron in Figs.~\ref{fig2} and \ref{fig3},
respectively.  In Fig~\ref{fig2}, we plot the $d\sigma/dQ_T$ of
$e^+e^-$ pairs as a function of $Q_T$ at $\sqrt{S}=1.8$~TeV.  The data
are from CDF Collaboration \cite{CDF-Z}.  Theory curves (Z-only) are 
from Eq.~(\ref{css-gen}) with the $\tilde{W}$ given in
Eqs.~(\ref{css-W-sol}) and (\ref{qz-W-sol}) for $b<b_{max}$ and
$b>b_{max}$, respectively.  CTEQ4 parton distribution and an overall
normalization 1.09 are used \cite{CDF-Z}.  In Fig.~\ref{fig3},    
we plot the $d\sigma/dQ_T$ for $W$ production with the same $b_{max}$
and $g_2$ and without any overall normalization \cite{QZ2}.  The data
for $W$ production are from D0 Collaboration \cite{D0-W}.  The QCD
predictions from the $b$-space resummation formalism are consistent
with the data for all $Q_T<Q$.  Furthermore, we let $g_2$ be a
fitting parameter for any given value of $b_{max}$.  Although the
fitting prefers $g_2\sim 0.4$~GeV$^{2}$, the $Q_T$ distributions are 
extremely insensitive to the choices of $b_{max}$ and $g_2$.  
The total $\chi^2$ is very stable for $b_{max} \in (0.25,
0.8)$~GeV$^{-1}$ and $g_2 \in (0, 1)$~GeV$^{2}$.  In Figs.~\ref{fig2}
and \ref{fig3}, we also plot the theory curves (dashed lines) with
$g_2=0.8$~GeV$^{2}$ (twice of the fitting value).  Non-vanish $g_2$
gives a small improvement to the $Q_T$ distributions at small $Q_T$.
We also vary the value of $\alpha$ in Eq.~(\ref{qz-fnp}) by
requiring only the first derivative to be continuous at $b=b_{max}$,
and find equally good theoretical predictions, except very mild
oscillations in the curves at very large $Q_T$ due to the Fourier
transform of a less smoother $b$-distribution.  The observed
insensitivity on $b_{max}$, $g_2$, and $\alpha$ is a clear 
evidence that the $b$-space resummation formalism is not sensitive to
the nonperturbative input at large $b$ for heavy boson production.  

We notice that the theory curve is below the data at large $Q_T$.  We
believe that it is because we have only the leading order contribution
to the $Y$ term in Eq.~(\ref{css-gen}).  At large $Q_T$, the $Y$ term
dominates.  Similar to the fixed-order perturbative calculations, the
next-to-leading order contribution will enhance the theoretical
predictions \cite{AER}.  In conclusion, the CSS $b$-space resummation
formalism has a good predictive power for heavy boson production at
Tevatron energy, and it should provide even better predictions at the
LHC energy \cite{QZ2}. 

We thank P. Nadolsky and C.P. Yuan for help on the Legacy program
package, and thank S. Kuhlmann for help on experimental data.  
This work was supported in part by the U.S. Department of Energy under
Grant No. DE-FG02-87ER40731.


\end{document}